\documentclass[12pt,twoside,a4paper]{article}
\usepackage[dvips]{epsfig}
\newcommand{\jet}{\rm jet}
\newcommand{\jets}{\rm jets}
\newcommand{\gev}{\rm GeV}
\begin{document}
\thispagestyle{empty} 
\title{
\vskip-3cm
{\baselineskip14pt
\centerline{\normalsize DESY 01--026 \hfill ISSN 0418--9833}
\centerline{\normalsize hep--ph/0103056 \hfill} 
\centerline{\normalsize March 2001 \hfill}} 
\vskip1.5cm
Photoproduction of Jets on a Virtual Pion Target \\
in Next-to-Leading Order QCD 
\author{M.~Klasen and G.~Kramer
\vspace{2mm} \\
{\normalsize II. Institut f\"ur Theoretische Physik, Universit\"at Hamburg,}\\ 
\normalsize{Luruper Chaussee 149, 22761 Hamburg, Germany} \vspace{2mm}
\\ 
} }

\date{}
\maketitle
\begin{abstract}
\medskip
\noindent
Differential cross sections for inclusive dijet photoproduction on a virtual
pion have been calculated in next-to-leading order QCD as a function of $E_T,
\eta$, and $x_{\pi}$. The cross sections are compared with recent ZEUS data
on photoproduction of dijets with a leading neutron in the final state.
\end{abstract}

\section{Introduction}

Recently the ZEUS collaboration at HERA presented differential cross section
data for the neutron tagged process $e^+ + p \rightarrow e^+ + n + \jet + \jet
+ X$ \cite{1}. The cross sections have been measured in the photoproduction
region with photon virtuality $Q^2 < 4~\gev^2$ for $\gamma p$ center-of-mass
energy $W$ in the interval $134 < W < 269~\gev$, for jets with transverse
energy $E_T > 6~\gev$, neutron energy $E_n > 400~\gev$, and neutron production
angle $\Theta_n < 0.8$ mrad. The cross sections were compared to predictions
with the one-pion exchange model and rather good agreement was found.\\

Due to the kinematic constraints on the neutron detection the squared momentum
transfer $t$ between ingoing proton and outgoing neutron is very small
\cite{1}. In this case it is expected that the $p \rightarrow n$ transition
amplitude, {\it i.e.} the amplitude for the production of neutrons, which carry
most of the momentum of the proton, is dominated by the lightest particle
in the $t$-channel, the pion. Hence the isolation of the one-pion exchange 
contribution provides the possibility to study the photon-pion interaction
$\gamma + \pi \rightarrow X$ \cite{2}.\\

Selecting high $E_T$ jets in the final state $X$ allows us to apply the QCD
improved parton model to describe the characteristic features of the jet
production dynamics. This is completely analogous to the process $\gamma + p
\rightarrow \jet + \jet + X$, which has been calculated in the past up to
next-to-leading order (NLO) using different computational techniques
\cite{3,4}. The only difference in $\gamma + \pi \rightarrow \jet+\jet+X$
is that the parton distribution functions (pdfs) of the proton are replaced
by the pdfs of the pion. In addition, the pion flux, generated by the 
$p \rightarrow n$ transition, has to be supplied.\\

The pion flux can, in principle, be measured in charge-exchange processes
in soft hadronic reactions, where an initial-state proton is transformed
into a final-state neutron, $p \rightarrow n$, with small momentum transfer.
A successful phenomenological description of the corresponding data has been
obtained in the framework of reggeized isovector exchange, such as $\pi,
\rho$, and $a_2$, with the pion being by far the lightest meson dominating
the $p \rightarrow n$ transition, particularly at small values of the
squared momentum transfer $t$ between proton and neutron \cite{5}.\\

The pdfs of the pion can be measured in deep-inelastic scattering with a
forward
going neutron detected  in the final state: $e^+ + p \rightarrow e^+ + n + X$
\cite{6}. The measurements of the structure functions so far are limited to 
rather small Bjorken-$x$ \cite{7,8}. The measured semi-inclusive cross sections
for leading neutrons by the H1 collaboration \cite{8} can be described entirely
by $\pi$ exchange, and the Bjorken-$x$ and $Q^2$ dependence of these data are
consistent with leading order (LO) pdfs of the pion. There exist several such
pion pdfs in the literature \cite{9,10,11}. They are constrained by dimuon and
prompt-photon production data from fixed target experiments that are sensitive
to the valence quark distribution in a Bjorken-$x$ range relevant for dijet
photoproduction on pions.\\

In this work we present the results of a calculation of differential cross
sections for the reaction $e^+ + p \rightarrow e^+ + n + \jet + \jet +X$ with
the kinematical constraints as in the ZEUS analysis \cite{1}. We have done 
these calculations in LO and in NLO. It is well known, however, that LO
predictions are not reliable due to the possibly strong scale dependence of the
results. Another problem of LO results for jet cross sections is the fact that
in LO only two large $E_T$ jets with opposite transverse momenta can be 
produced. Therefore these cross sections are independent of the jet algorithm 
which is applied in order to define jets in the analysis of the experimental 
data. In NLO calculations there appears an explicit dependence on the jet 
algorithm due to the possible production of three partons (in addition to the
remnant partons on the photon or pion side, respectively) in the final state,
two of which must be combined in one jet in accordance with the 
experimental constraints for the construction of jets out of hadrons.
Therefore our LO results are just for comparison and for demonstrating the 
reduced scale dependence of the NLO results. Our genuine predictions are the 
NLO cross sections which will be compared with the ZEUS data \cite{1}.\\

In the next section we shall describe the kinematical variables and define our
input for the pion flux and the pion pdfs. Section 3 contains our results and
the comparison with the experimental data. In the last section we give a short
summary and draw some conclusions.\\ 

\section{Kinematical Variables and Further Input}

The event kinematics has been described in detail already in \cite{1}. Here we
repeat those variables which are needed in the calculation of the cross 
cross sections. The reaction 
\begin{equation}
  e^+(k) + p(P) \rightarrow e^+(k') + n(P') +H,
\end{equation}
where H is the hadronic system containing the jets, is characterized by the
four-momenta $k$ and $k'$ of the initial and scattered positron and by $P$
and $P'$, the four-momenta of the ingoing proton and outgoing neutron, 
respectively. The positron-photon vertex is described by the exchanged photon
virtuality $Q^2$ and the positron's inelasticity $y$, {\it i.e.}
\begin{eqnarray}
    Q^2 &=& -q^2 = -(k-k')^2, \nonumber \\
    y   &=& \frac{Pq}{Pk}.
\end{eqnarray}
In the ZEUS experiment \cite{1} $Q^2 \leq 4~\gev^2$ and $0.2<y<0.8$,
corresponding to the $\gamma p$ center-of-momentum energy range
$134 < W < 269~\gev$. The protons at HERA have the energy $E_p = 820~\gev$
and collide with $E_e = 27.5~\gev$ positrons, corresponding to a 
center-of-momentum frame energy $\sqrt{S} = 300~\gev$. The two variables, which
describe the proton-neutron vertex, are the fraction of the energy of the 
initial-state proton carried by the neutron $x_L$ and the square of the
momentum transfer $t$ between the proton and the produced neutron:
\begin{eqnarray}
            x_L &=& \frac{P'k}{Pk} \simeq \frac{E_n}{E_p}, \nonumber \\
            t   &=& q'^2 =(P-P')^2.
\end{eqnarray}

The splitting function or pion flux for the transition $p \rightarrow n+\pi^+$
is usually parameterized by different forms which can be summarized as
\begin{equation}
    f_{\pi/p}(x_L,t)= \frac{1}{4\pi}\frac{g_{n\pi p}^2}{4\pi}
    \frac{-t}{(m_{\pi}^2-t)^2}(1-x_L)^{1-2\alpha_{\pi}(t)}[F(x_L,t)]^2.
\end{equation}
Here $g_{n\pi p}$ is the coupling constant of the $n\pi p$ vertex, $m_{\pi}$ 
is the pion mass, and $\alpha_{\pi}(t)=\alpha'(t-m_{\pi}^2)$ is the Regge 
trajectory of the pion. $F(x_L,t)$ is a form factor which describes the 
off-shell behavior of the virtual pion and/or possible final state 
rescattering effects of the neutron. Two choices for the form factor 
$F(x_L,t)$ used in the analysis of various charge-exchange scattering 
experiments are
\begin{equation}
  F(x_L,t) = \left\{ \begin{array} {l} \exp [b(t-m_{\pi}^2)]  \\ 
                                       \exp [R^2(t-m_{\pi}^2)/(1-x_L)] 
                     \end{array} \right.
\end{equation}
where $b$ and $R$ are constants. The first of these choices, the exponential
form, is usually taken with the Regge trajectory factor in (4) with 
$\alpha ' = 1~\gev^{-2}$. The second choice, the light-cone form factor, is 
usually associated with the flux without the Regge trajectory factor,
{\it i.e.}
$\alpha ' = 0$ in (4). These choices are associated with the experimental 
finding that in soft hadronic reactions the shape of the $x_L$ distribution
depends on $t$ \cite{12}. The pion-nucleon coupling constant is well known
from low energy $\pi N$ and $NN$ scattering data. We take 
$g_{n\pi p}^2/4\pi =2\cdot 14.17$ as obtained in a recent analysis \cite{13}.\\

In the scattering process of $2 \rightarrow 2$ massless partons, the fractions
of the four-momenta $q=k-k'$ and $q'=P-P'$ participating in the hard scattering
by the initial state parton are given by
\begin{eqnarray}
 x_{\gamma} &=& \frac{\sum_{j} E_T^j e^{-\eta_j}}{2yE_e},     \\
 x_{\pi}    &=& \frac{\sum_{j} E_T^j e^{\eta_j}}{2E_p(1-x_L)},
\end{eqnarray}
where the sums in (6) and (7) run over the variables of the two jets in the 
final state. Here we assumed that $q^2=q'^2=0$. The energy fraction 
contributing by the exchanged virtual photon to the production  
of the dijets is $x_{\gamma}$ whereas the corresponding contribution
of the virtual pion (or possibly of a reggeized $\rho$ or $a_2$) is $x_{\pi}$.
In (6) $E_{\gamma}=yE_e$ is the energy of the ingoing virtual photon.
In NLO also three jets can be produced in the final state. Then (6) and (7)
are no longer valid. In order to estimate the energy fractions $x_{\gamma}$
and $x_\pi$ for this case, too, one uses (6) and (7) in the form that the
sums in (6) and (7) run over the two jets of largest $E_T$ in an event.\\

As is well known two mechanisms contribute to the photoproduction of jets,
the direct and the resolved process. The observable $x_{\gamma}$ is sensitive
to the amount of direct and resolved processes. The LO direct process, where
the photon couples directly to the quarks, contributes at $x_{\gamma}=1$, the 
resolved and the NLO direct processes contribute in the region $x_{\gamma}<1$.
It must be emphasized, however, that the distinction of direct and 
resolved processes looses its meaning in NLO, so that the characterization of
the contributions to the various $x_{\gamma}$ regions by the direct and 
resolved processes is not unique: It becomes scale dependent.
The variable $x_{\pi}$ can be used to analyze which region of the pion pdfs and
which partonic contributions are most important for dijet production with
leading neutrons.\\

As for the jet definition and the combination of two partons into one jet
in the NLO contributions of the direct and the resolved parts,
we use the $k_T$ cluster algorithm
in the form as introduced in \cite{14} and as used in the jet analysis of the
experimental data \cite{1}.
The photon flux is calculated with the usual Weizs\"acker-Williams
approximation, including the non-logarithmic correction as calculated in
\cite{14a}. For the resolved cross sections
we need the pdfs of the photon in LO and NLO. A popular parameterization is GRV
\cite{15}. We use the parameterization GS96 \cite{16}, since it leads to a 
slightly better description of the experimental data. For the comparison with 
data of inclusive jet production without the leading neutron, which has been 
simultaneously analyzed in \cite{1} under similar kinematical conditions, we 
need also the pdfs of the proton, for which we employ the recent CTEQ5M
\cite{17} parameterization. The $\Lambda $ parameter which we need in the
NLO $\alpha_s$ formula is taken from the proton pdf fit. It is $\Lambda^{(4)}
=326$ MeV. For the pion pdfs we employ two alternatives, SMRS3 \cite{9} and
GRS \cite{11}. Unfortunately, the $\Lambda $ values used for the evolution of
these pdfs are somewhat smaller, namely $\Lambda^{(4)} = 190$ and 200 MeV.
For LO predictions, we use LO matrix elements with the one-loop formula for
$\alpha_s$ and the same values of $\Lambda$. Since only GRS have constructed
LO and NLO pion pdfs, we have to use the NLO SMRS3 set also with our LO
predictions. The GRS pdfs of the pion are constructed for three flavors only.
Since we include four flavors throughout, we also need the pion pdf for charm.
This is taken from the earlier work of GRV \cite{10}.\\

\section{ Results and Comparison with ZEUS Data}

The calculation of the cross sections is based on the formalism fully described
in our previous work \cite{3}. To cancel infrared and collinear singularities
in the NLO contributions of the final state and to define the collinear initial
state singularities to be removed through the renormalization of the
pdfs of the photon or the pion, the phase-space slicing method was applied.
Since the neutron kinematics could not be fixed in detail in the experiment
we had to integrate over a finite region in $x_L$ and $t$. We did this in
accordance with the specifications of the ZEUS experimental analysis.
Except for the outgoing positron and the leading neutron the final state 
consists of two or three jets. The two-jet sample contains the bare parton
jets from the LO and virtual NLO contributions and the two jets originating
from the recombination of two partons in the three-parton terms. The three-jet
sample is just given by the uncombined three-parton final states.\\ 

First we calculated the differential cross section $d^2\sigma/dE_Td\eta$,
where $E_T$ and $\eta $ are the transverse energy and the rapidity of one
of the jets in the two- or three-jet sample, respectively. As in the ZEUS
analysis we restricted the selection in the three-jet sample. We included
from this sample only the two jets with the largest $E_T$. In our earlier
work \cite{3} this cross section was denoted the inclusive one-jet cross
section. There, however,  we included all three jets from the three-jet
contributions. In \cite{1} the cross section defined above is denoted
the dijet cross section. This should be distinguished from the inclusive
two-jet cross section defined in \cite{3}, which depends on $E_T$, the
transverse energy of one of the two jets with the highest $E_T$, and the
rapidities $\eta_1$ and $\eta_2$ of these high $E_T$ jets. The $d^2\sigma/
dE_Td\eta$ is integrated over $-2<\eta<2$ as in \cite{1}. The result for the
full jet sample (without detection of the leading neutron) as a function of
$E_T$ for $E_T > 6~\gev$ turn out to be in reasonable agreement with the 
experimental data from \cite{1} (not shown). The point with the largest
cross section at $E_T = 7~\gev$ agrees perfectly with the measured value. The 
points at medium $E_T$ lie somewhat higher than the theoretical curve. 
The equivalent cross section $d^2\sigma/dE_Td\eta$, also integrated over 
$|\eta| < 2$, for the case with the leading neutron is shown as a function of
$E_T$ for $6.5<E_T<29.5~\gev$ in Fig. 1a.
\begin{figure}
 \begin{center}
  \vspace*{-2cm}
  \epsfig{file=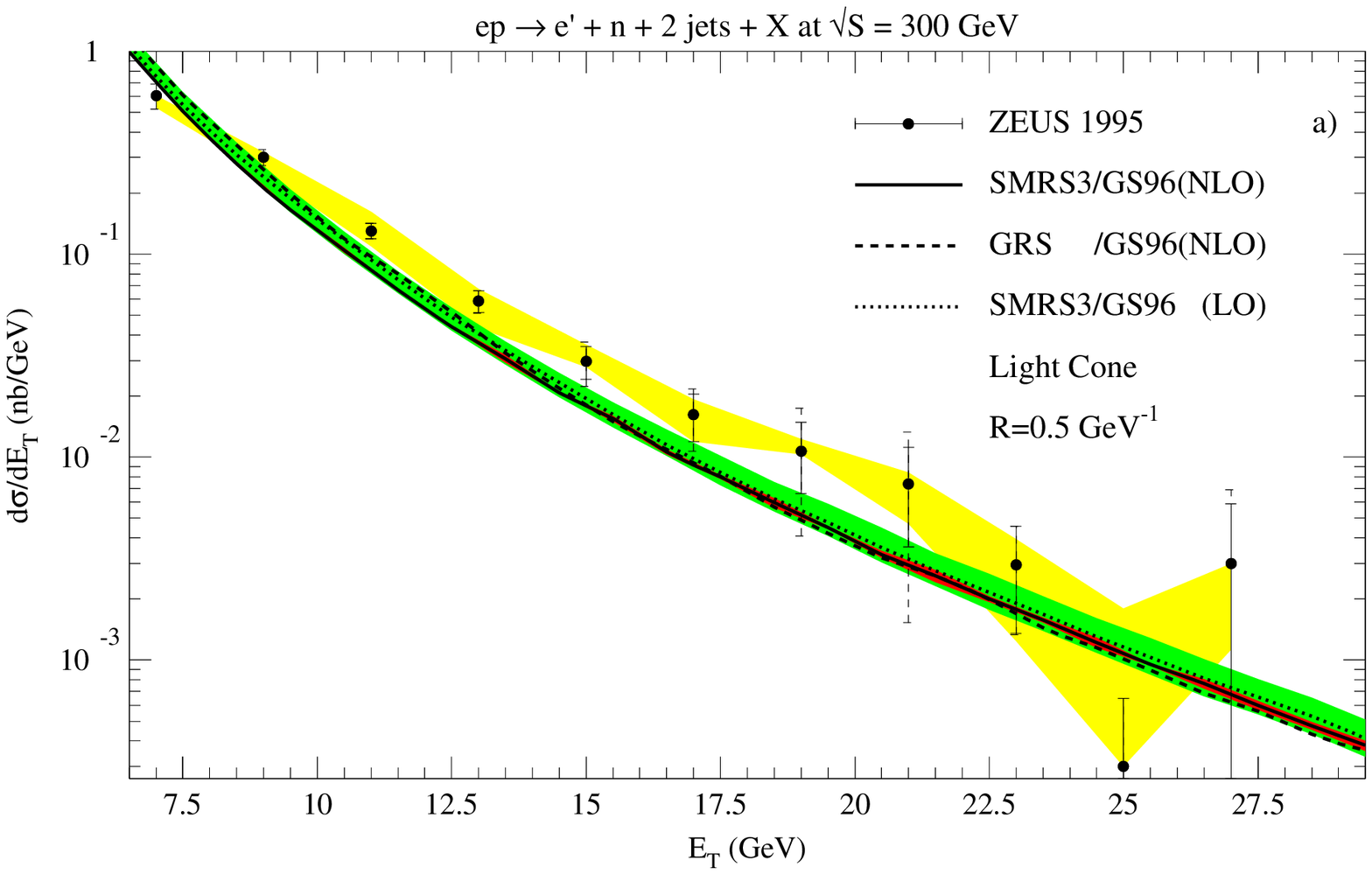,width=\textwidth}
 \end{center}
 \begin{center}
  \vspace*{-2cm}
  \epsfig{file=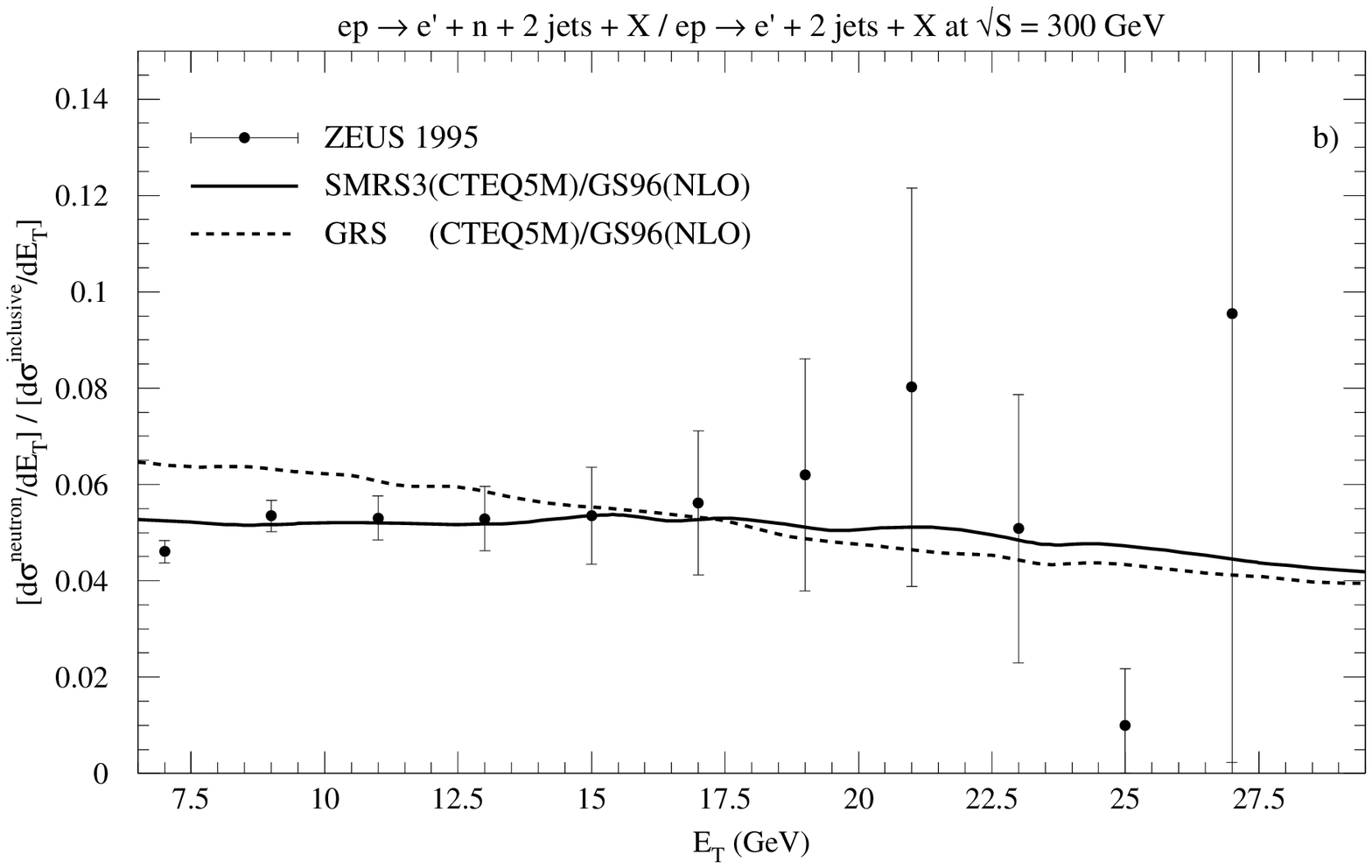,width=\textwidth}
 \end{center}
 \vspace*{-1.3cm}
 \caption{\label{fig:1}
 Dependence of the dijet photoproduction cross section with a leading neutron
 on the transverse energy of one of the two jets (a). Two NLO predictions with
 different pion structure functions and one LO prediction are shown together
 with experimental data from ZEUS with statistical (inner) and systematic
 (outer) experimental error bars. In
 addition, error bands from the theoretical scale uncertainty (medium: LO,
 dark: NLO) and the experimental energy scale uncertainty (light) are shown.
 The NLO error band coincides with the thickness of the full line.
 In b) we present the ratio of the leading neutron over the inclusive
 cross section versus the transverse energy. NLO calculations with two
 different pion structure functions are shown together with the experimental
 ZEUS data. In this case, only statistical error bars are given.}
\end{figure}
In this figure three curves are shown:
one from LO with pion pdfs SMRS3 \cite{9} and photon pdfs GS96 \cite{16}, the
latter
in LO, and two other curves for NLO with pion pdfs SMRS3 \cite{9} and 
GRS \cite{11} and charm content from
GRV \cite{10} for comparison. We see that all three predictions lie very near
to each other in the logarithmic plot. The LO and NLO curves for SMRS3/GS96
differ only very little in this plot, showing that the NLO corrections are 
small for the scale choice $\mu = \max(E_{T,1},E_{T,2})$
as used in Fig. 1a (and in all the
following figures) for the LO and NLO result. The GRS choice for the pion
pdfs leads to a somewhat larger cross section at small $E_T$'s and smaller 
cross section at the larger $E_T$'s as compared to the SMRS3 choice.
We have calculated that the gluon in the pion gives the dominant contribution
to the cross section for $E_T<22$ GeV. Resolved photons are more important
than direct photons for $E_T<9$ GeV.
The
experimental data \cite{1} with statistical and systematic errors shown
separately come with a light shaded band which indicates the additional
systematic experimental uncertainty due to the calorimeter energy scale
\cite{1}. The medium shaded band gives the uncertainty of the LO result when
the scale of $\alpha_s$ and of the factorization is varied in the interval
between $E_T/2$ and $2\cdot E_T$. The scale uncertainty of the NLO prediction
(dark shaded band) coincides with the line thickness of the NLO curve.
Taking into account the fairly large 
experimental errors including the calorimeter energy scale uncertainty the 
agreement of the ZEUS data with the theoretical predictions concerning shape 
and absolute normalization is reasonably good. It must be emphasized, however, 
that the absolute normalization depends on the pion form factor (5). In 
Fig. 1a we assumed the light-cone form factor with $R=0.5~\gev^{-1}$. Recent
determinations of $R^2$ from a comparison to the rather accurate neutron 
production data from Flauger and M\"onning \cite{12} yield $R^2=0.2~\gev^{-2}$
\cite{5} and $R^2=0.4~\gev^{-2}$ \cite{17a}
in reasonable agreement with our value for $R^2$. A change of $R$
from $R=0.5$ to $R=0.6~\gev^{-1}$, {\it i.e.}
an increase by $20\%$, leads to a  $15\%$ smaller cross section.\\

Since we have calculated also $d^2\sigma/dE_Td\eta$ with the same kinematical
constraints for $e+p \rightarrow e' + 2~\jets +X$ we can take the ratio of the
two cross sections $d\sigma/dE_T$ for $e+p \rightarrow e'+n+2~\jets +X$ and
$e+p \rightarrow e'+ 2~\jets +X$. This is shown in Fig. 1b and compared to the
ZEUS data for this ratio. The agreement with the data is perfect. For low 
$E_T$, where the experimental errors are smallest, the SMRS3
result agrees better than the ratio obtained with the GRS pion pdfs, 
otherwise  the results are the same. In Fig. 1b the experimental uncertainty
from the calorimeter energy measurements, the systematic error and the scale
uncertainty in the theoretical curve are not included since they are supposed 
to cancel in the ratio \cite{1}. Thus, this ratio seems to be a good example to
get information on the pion pdfs and/or the pion form factor.\\

The next calculated differential cross section is $d\sigma/d\eta$, {\it i.e.}
$d^2\sigma/dE_Td\eta$ integrated over $E_T$ with $E_{T,\min} = 6~\gev$. The
result is shown in Fig. 2a,
\begin{figure}
 \begin{center}
  \vspace*{-2cm}
  \epsfig{file=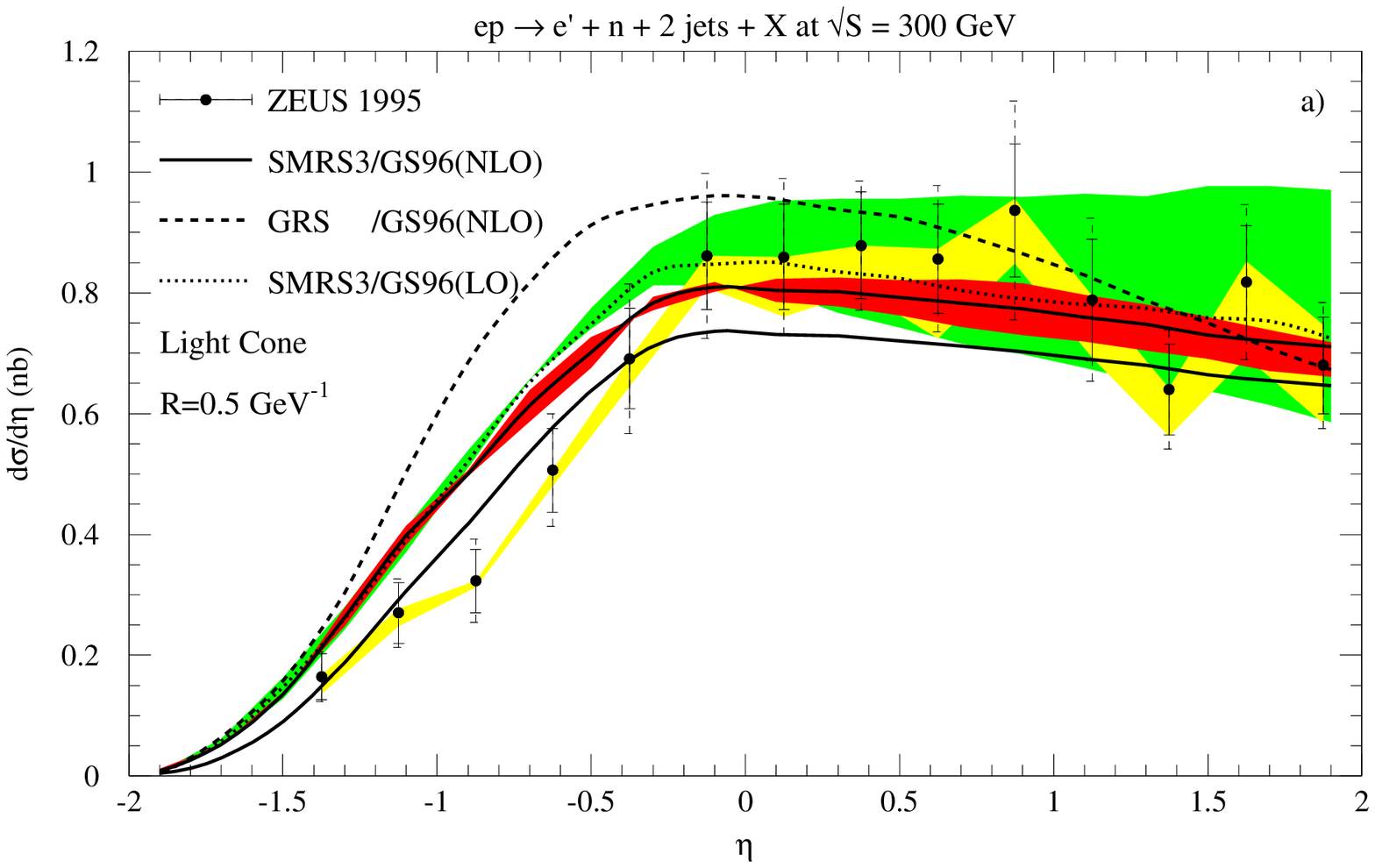,width=\textwidth}
 \end{center}
 \begin{center}
  \vspace*{-2cm}
  \epsfig{file=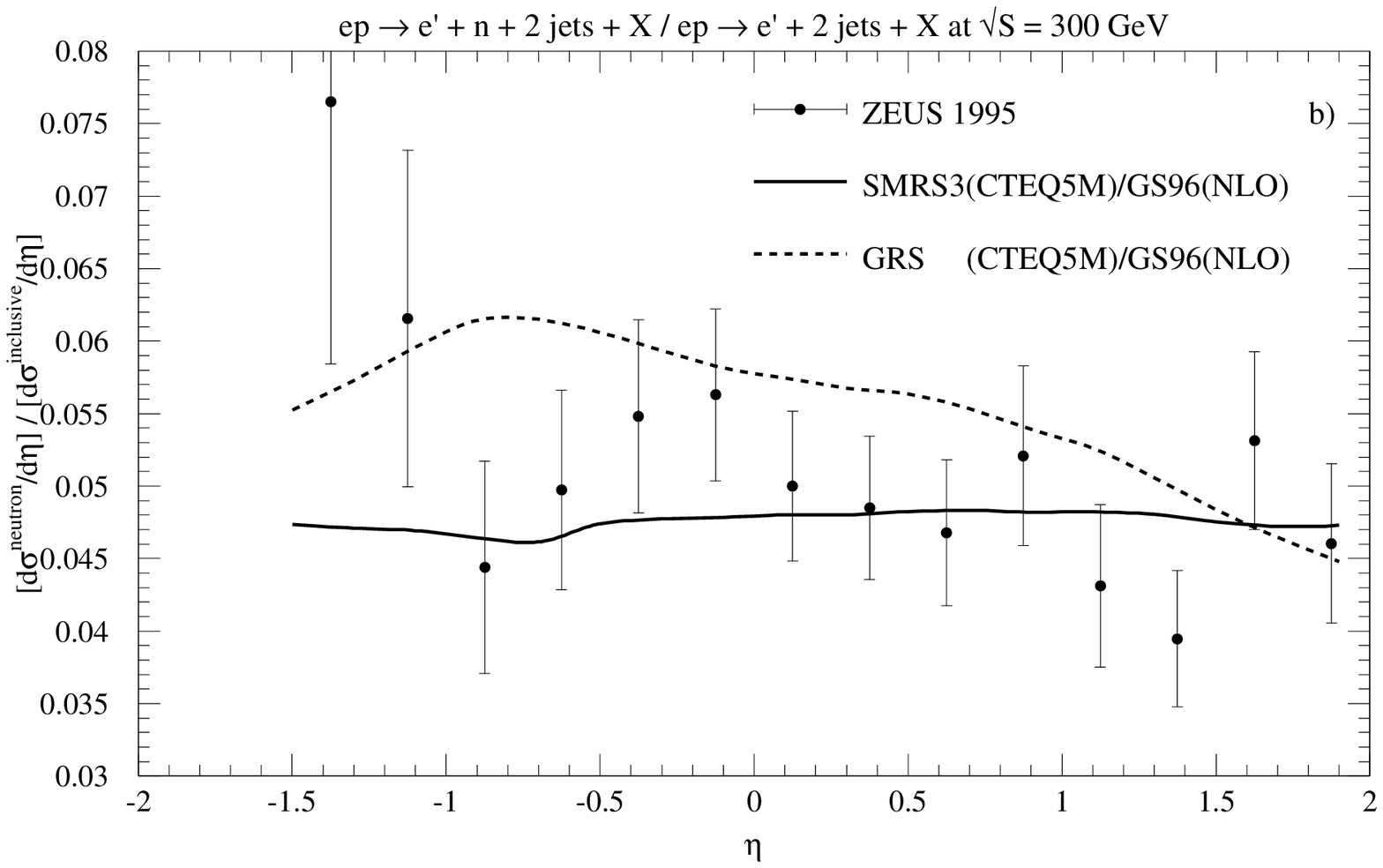,width=\textwidth}
 \end{center}
 \vspace*{-1cm}
 \caption{\label{fig:2}
 Dependence of the dijet photoproduction cross section with a leading neutron
 on the rapidity of one of the two jets (a). In b) we show the ratio of the
 leading neutron over the inclusive cross section versus the rapidity. Details
 are as in Fig.\ 1. The lower full curve in a) demonstrates the influence of
 hadronization corrections.}
\end{figure}
where $d\sigma/d\eta$ is plotted as a function of
$\eta$ for $\eta$ between $-2$ and $2$, together with the data points from
\cite{1}. There are three theoretical curves for the three cases as in 
Fig. 1a. Here one can see the difference between SMRS3/GS96 in LO and NLO.
The difference is very small, the LO cross section is approximately $5\%$
larger than the NLO cross section for the same scale
$\mu = \max(E_{T,1},E_{T,2})$. Concerning
the dependence on the pion pdfs the choice SMRS3 agrees best with the data. The
result with the GRS pdfs is larger, in particular for negative $\eta$.
In \cite{9} two other pion pdfs have been published, SMRS1 and SMRS2. The
cross sections for these two pdfs lie between the SMRS3 and the
GRS curves and are therefore not shown explicitly.
The gluons in the pion contribute about twice as much as quarks to the
rapidity distribution over the entire $\eta$-range. Direct (resolved)
photons dominate in the region $\eta<0$ $(\eta>0)$.
All three curves
do not agree well with the data points for $\eta < -0.5$. For jet production
without leading neutron this region is known to be influenced by hadronization
corrections \cite{18}. We expect such corrections also for the cross section in
Fig. 2a, although somewhat less in the case with leading neutron due to 
the reduced total energy producing the hard jets. The theoretical result with
such hadronic corrections included (taken from the case without leading neutron
\cite{18}) is shown in Fig. 2a as the lower full curve. It leads to a better
agreement in the negative $\eta$ region, but it produces also a somewhat lower
cross section at positive $\eta$'s, reducing the good agreement wit the data
in this region. In Fig. 2a we show again three shaded bands: (i) the light 
shaded band shows the additional systematic uncertainty from the calorimeter
energy scale, (ii) the medium shaded band gives the variation of the
LO result (dotted curve) with the scale $\mu$, which was varied as usual in
the region $E_T/2 < \mu < 2\cdot E_T$, (iii) the dark shaded band shows the
variation of the NLO cross section with changing the scale in the same 
interval. Here, we see quite clearly that the NLO result has a very much 
reduced scale dependence compared to the LO result. Similarly to Fig. 1b we
have plotted in Fig. 2b the ratio of the cross sections for SMRS3/GS96 and
GRS/GS96 and the cross section $d\sigma/d\eta$ for dijet production without
leading neutron as a function of $\eta$ and compared them with the same
experimental ratio from ZEUS. Within errors the SMRS3 and the GRS results are
consistent with the data, the SMRS3 curve agrees slightly better.\\

The last differential cross section which we investigated is $d\sigma/
d\log_{10}(x_{\pi})$ as a function of $\log_{10}(x_{\pi})$.
In order to determine $x_{\pi}$, which is a measure
of the exchanged pion's momentum fraction participating in the production of
jets as defined in (7), we need the transverse momenta and rapidities of the
two most energetic jets in the event. The experimental cross section 
$d\sigma/d\log_{10}(x_{\pi})$, however, has the constraint that for the two
jets $E_{T,1} \geq 6~\gev$ {\em and}
$E_{T,2} \geq 6~\gev$. Cross sections for $E_{T,1}=E_{T,2}$ are problematic
theoretically, since they become infrared sensitive in NLO,  
{\it i.e.} depend on the slicing cut used to cancel infrared and collinear 
singularities. In order to avoid this sensitivity and also possible hadronic
and intrinsic $k_T$ effects not considered in our NLO framework we need
constraints on $E_{T,1}, E_{T,2}$ or $E_{T,3}$ which smear out the problematic
region $E_{T,1}=E_{T,2}$. This problem is well known and was encountered some 
time ago also in connection with the calculation of the inclusive dijet cross 
section in $\gamma p$ collisions \cite{19}. One remedy to remove the infrared
sensitivity is the requirement of slightly different lower limits on $E_{T,1}$
and $E_{T,2}$. Therefore, we have calculated $d\sigma/d\log_{10}
(x_{\pi})$ with the constraint
$E_{T,1} \geq 6$ GeV, $E_{T,2} \geq 5$ GeV if $E_{T,1} > E_{T,2}$ or
$E_{T,1}$ and $E_{T,2}$ interchanged if $E_{T,2} > E_{T,1}$. The cross sections
calculated with this constraint are plotted in Fig. 3, again for the three 
cases as in Fig. 1a.
\begin{figure}
 \begin{center}
  \epsfig{file=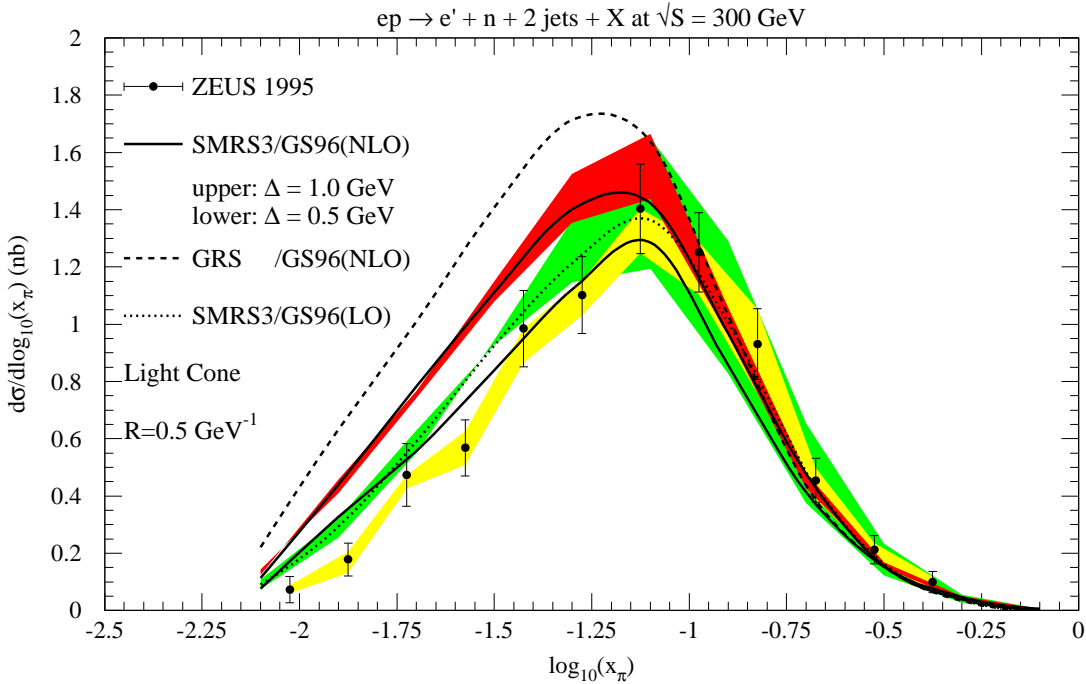,width=\textwidth}
 \end{center}
 \caption{\label{fig:3}
 Dependence of the dijet photoproduction cross section with a leading neutron
 on the logarithm of the observed momentum fraction of the partons in the pion
 $\log_{10}(x_\pi)$. Details are as in Figs.\ 1a and 2a. In addition, we show
 the effect of lowering the transverse energy difference $\Delta$ from 1 GeV
 to 0.5 GeV.}
\end{figure}
These cross sections are compared with the corresponding ZEUS data.
The NLO prediction with the pion pdfs of SMRS3 agrees better with the data
than the one with GRS pion pdfs. Except for large $x_\pi > 0.35$ the gluons
in the pion dominate over quarks, and the resolved photon contribution is
larger than the direct one for $x_\pi > 0.04$.
The agreement with the data is good for $x_{\pi} > 0.06$,
but less good for the smaller $x_{\pi}$. This $x_{\pi}$ region is related to 
the negative $\eta$ region in the $\eta$ distribution in Fig. 2a, where one has
hadronization corrections (see (7) for the relation between $x_{\pi}$ and the
$\eta_j$'s). We note that the K factor between the NLO and LO result is 
larger than in $d\sigma/d\eta$ and is larger than one now in particular for
small $x_{\pi}$. This is connected with the specific constraint on $E_{T,1}$
and $E_{T,2}$, which was not necessary for the results in Figs.~1 and 2.
This constraint does not exactly 
correspond to the one applied to the measured $d\sigma/d\log_{10}
(x_{\pi})$. On 
the other hand the low $E_T$ calorimeter jets used in the experimental analysis
have an uncertainty of $5\%$ which introduces a corresponding uncertainty on
the constraint $E_{T,1}, E_{T,2} \geq 6~\gev$. In order to see the influence of
the $E_{T,1},E_{T,2}$ cut we reduced the difference $\Delta$
between $E_{T,1}$ and
$E_{T,2}$ from $1~\gev$ to $0.5~\gev$. The result is shown in Fig. 3 as the
lower full curve. The reduction of the difference leads to a lower cross
section as we expect, so that the low $x_{\pi}$ experimental points are
described very well now but not the higher $x_{\pi}$ region. Since this region
is much less affected by hadronization corrections we prefer the original
prediction (upper full curve in Fig. 3). The absolute normalization of the
cross section $d\sigma/d\log_{10}(x_{\pi})$
depends also on the pion form factor. The
curves in Fig. 3 are calculated with the light-cone form factor with $R=0.5~
\gev^{-1}$. $R=0.6~\gev^{-1}$
instead gives a $15\%$ lower cross section and better agreement for 
$x_{\pi} < 0.06$ but worse agreement above this value. Again taking into 
account possible hadronization corrections we consider the value
$R=0.5~\gev^{-1}$ as more realistic. The result with $R=0.6~\gev^{-1}$
is almost indistinguishable from
the result with the exponential form factor with $b=0$.\\

In \cite{1} several uncorrected distributions are presented, in particular
event rates as a function of $x_{\gamma}$ in order to say something about
the amount of resolved and direct photoproduction. From these rates, the
ratio of resolved (corresponds to $x_{\gamma} < 0.75$) to the direct cross
section (corresponds to $x_{\gamma} > 0.75$) has been deduced.
We checked that the theoretical prediction
for this ratio is in reasonable agreement with the data. It decreases with
increasing $E_T$, as one expects it. \\

In this work we considered only the pion exchange in the $t$-channel. Other
isovector exchanges, such as $\rho$ and $a_2$, are expected to be small.
Recent studies of leading neutron production in $ep$ collisions show that
the rate can be described entirely by $\pi^+$ exchange \cite{8}. The additional
exchanges, which increase the rate of neutron production, are offset by
absorptive rescattering of the neutron, which decreases the rate by
approximately the same amount \cite{5}. Therefore both effects are neglected
in the present analysis. If the experimental data were more specific
concerning the dependence on $t$ and $x_L$, the dominance of the $\pi^+$
exchange could be demonstrated.

\section{Summary and Concluding Remarks} 

We have presented a next-to-leading order calculation of dijet production
in collisions of almost real photons and pions, where the photons are generated
through bremsstrahlung from an initial electron or positron beam and the pion
is assumed to give the dominant contribution to the forward proton to neutron
transition amplitude. Our results have been compared with recent data from ZEUS
for distributions in the transverse energy $E_T$ and rapidity $\eta$ of one of
the two jets and for the observed momentum fraction of the partons in the pion
$x_\pi$. We found that the NLO corrections to the LO result were generally
small, also in the case of inclusive photon-proton scattering with similar
kinematic constraints, but lead to a largely reduced dependence on the
renormalization and factorization scales. For the NLO
distribution in $\log_{10}(x_\pi)$, we had to relax the experimental condition
of $E_{T,1},E_{T,2} > 6$ GeV to allow for a small difference $\Delta=0.5...1$
GeV of the two minimal transverse energies, corresponding to the non-zero
experimental uncertainty in the jet energy scale. \\

The overall normalization of all three distributions is sensitive to the form
factor of the $p\rightarrow n\pi^+$ transition. We have used a light-cone
form with a value for the parameter $R$ which agrees well with recent
determinations from neutron production data. Once the normalization is fixed,
the shapes of the distributions, particularly the one in $\log_{10}(x_\pi)$,
are sensitive to the parton distributions in the pion.
The ZEUS data for the distribution in $\log_{10}(x_\pi)$ as well as in $E_T$
and $\eta$ are best described by the SMRS3 pion parameterization, whereas the
GRS parameterization with GRV charm distribution gives larger predictions,
especially at small $x_\pi$. We emphasize that for small $E_T$ one is
particularly sensitive to the gluon content of the pion.
Furthermore, the GS96 photon densities agree
better with the data than GRV. Hadronization corrections were found
to reduce the perturbative prediction considerably in the region of small
$\eta$, which corresponds to small $x_\pi$.

\subsection*{Acknowledgments}

We thank M.~Khakzad, G.~Levman and J.~Whitmore for making the ZEUS data
available to
us and M.~Wing for calculating the hadronization corrections with the HERWIG
Monte Carlo generator. Financial support by the
Deutsche Forschungsgemeinschaft through Grant No.\ KL~1266/1-1, by the
Bundesministerium f\"ur Bildung und Forschung through Grant No.\ 05~HT9GUA~3,
and by the European Commission through the Research Training Network
{\it Quantum Chromodynamics and the Deep Structure of Elementary Particles}
under Contract No.\ ERBFMRX-CT98-0194 is gratefully acknowledged.



\begin{thebibliography}{99}
  
\bibitem{1}
J.~Breitweg {\it et al.}  [ZEUS Collaboration],
DESY-00-142, October 2000, hep-ex/0010019.
 
\bibitem{2}
G.~Kramer, Proc.\ of the Seminar on {\it e--p and e--e Storage Rings}, ed.\ by
J.K.~Bienlein, I.~Dammann and H.~Wiedemann, DESY-73-66, December 1973.

\bibitem{3}
M.~Klasen and G.~Kramer,
Z.\ Phys.\ C {\bf 76} (1997) 67; \\
M.~Klasen, T.~Kleinwort and G.~Kramer,
Eur.\ Phys.\ J.\ direct C {\bf 1} (1998) 1.

\bibitem{4}
B.~W.~Harris and J.~F.~Owens,
Phys.\ Rev.\ D {\bf 56} (1997) 4007; \\
S.~Frixione and G.~Ridolfi,
Nucl.\ Phys.\ B {\bf 507} (1997) 315; \\
P.~Aurenche, L.~Bourhis, M.~Fontannaz and J.~P.~Guillet,
Eur.\ Phys.\ J.\ C {\bf 17} (2000) 413.

\bibitem{5}
U.~D'Alesio and H.~J.~Pirner,
Eur.\ Phys.\ J.\ A {\bf 7} (2000) 109
and the earlier literature given there.

\bibitem{6}
J.~D.~Sullivan, Phys.\ Rev.\ D {\bf 5} (1972) 1732; \\
N.~S.~Craigie and G.~Schierholz,
Nucl.\ Phys.\ B {\bf 100} (1975) 125.

\bibitem{7}
M.~Derrick {\it et al.}  [ZEUS Collaboration],
Phys.\ Lett.\ B {\bf 384} (1996) 388.

\bibitem{8}
C.~Adloff {\it et al.}  [H1 Collaboration],
Eur.\ Phys.\ J.\ C {\bf 6} (1999) 587.

\bibitem{9}
P.~J.~Sutton, A.~D.~Martin, R.~G.~Roberts and W.~J.~Stirling,
Phys.\ Rev.\ D {\bf 45} (1992) 2349.

\bibitem{10}
J.~F.~Owens,
Phys.\ Rev.\ D {\bf 30} (1984) 943; \\
P.~Aurenche, R.~Baier, M.~Fontannaz, M.~N.~Kienzle-Focacci and M.~Werlen,
Phys.\ Lett.\ B {\bf 233} (1989) 517; \\
M.~Gl\"uck, E.~Reya and A.~Vogt,
Z.\ Phys.\ C {\bf 53} (1992) 651.

\bibitem{11}
M.~Gl\"uck, E.~Reya and I.~Schienbein,
Eur.\ Phys.\ J.\ C {\bf 10} (1999) 313.

\bibitem{12}
J.~Engler {\it et al.},
Nucl.\ Phys.\ B {\bf 84} (1975) 70; \\
W.~Flauger and F.~M\"onnig,
Nucl.\ Phys.\ B {\bf 109} (1976) 347; \\
B.~Robinson {\it et al.},
Phys.\ Rev.\ Lett.\ {\bf 34} (1975) 1475; \\
G.~Levman and K.~Furutani,
DESY-95-142, July 1995, \\
and further references given there.

\bibitem{13}
T.~E.~Ericson, B.~Loiseau and A.~W.~Thomas,
hep-ph/0009312.

\bibitem{14}
S.~Catani, Y.~L.~Dokshitzer, M.~H.~Seymour and B.~R.~Webber,
Nucl.\ Phys.\ B {\bf 406} (1993) 187; \\
S.~D.~Ellis and D.~E.~Soper,
Phys.\ Rev.\ D {\bf 48} (1993) 3160.

\bibitem{14a}
S.~Frixione, M.~L.~Mangano, P.~Nason and G.~Ridolfi,
Phys.\ Lett.\ B {\bf 319} (1993) 339.

\bibitem{15}
M.~Gl\"uck, E.~Reya and A.~Vogt,
Phys.\ Rev.\ D {\bf 46} (1992) 1973.

\bibitem{16}
L.~E.~Gordon and J.~K.~Storrow,
Nucl.\ Phys.\ B {\bf 489} (1997) 405.

\bibitem{17}
H.~L.~Lai {\it et al.}  [CTEQ Collaboration],
Eur.\ Phys.\ J.\ C {\bf 12} (2000) 375.

\bibitem{17a}
H.~Holtmann, A.~Szczurek and J.~Speth,
Nucl.\ Phys.\ A {\bf 596} (1996) 631.

\bibitem{18}
B.~W.~Harris, M.~Klasen and J.~Vossebeld,
Proc.\ of the Workshop on {\it Monte Carlo Generators at HERA Physics},
Hamburg, Germany, 1998, hep-ph/9905348; \\
M.~Wing, priv.\ communication.

\bibitem{19}
M.~Klasen and G.~Kramer,
Phys.\ Lett.\ B {\bf 366} (1996) 385.

\end{thebibliography}
\end{document}